\documentclass[conference]{IEEEtran}
\IEEEoverridecommandlockouts

\usepackage{CJKutf8}
\usepackage{cite}
\usepackage{amsmath,amssymb,amsfonts}
\usepackage{algorithmic}
\usepackage{graphicx}
\usepackage{textcomp}
\usepackage{xcolor}
\def\BibTeX{{\rm B\kern-.05em{\sc i\kern-.025em b}\kern-.08em
    T\kern-.1667em\lower.7ex\hbox{E}\kern-.125emX}}
\usepackage{subfigure}

\begin{document}

\title{Memristive oscillatory circuits for resolution of NP-complete logic puzzles: Sudoku case}

\author{\IEEEauthorblockN{Theodoros Panagiotis Chatzinikolaou\IEEEauthorrefmark{1},
Iosif-Angelos Fyrigos\IEEEauthorrefmark{1},
Rafailia-Eleni Karamani\IEEEauthorrefmark{1},
Vasileios Ntinas\IEEEauthorrefmark{1}\IEEEauthorrefmark{2},\\
Giorgos Dimitrakopoulos\IEEEauthorrefmark{1},
Sorin Cotofana\IEEEauthorrefmark{3}, 
Georgios Ch. Sirakoulis\IEEEauthorrefmark{1}, }
\IEEEauthorblockA{\IEEEauthorrefmark{1}Department of Electrical and Computer Engineering, Democritus University of Thrace, Xanthi, Greece}
\IEEEauthorblockA{\IEEEauthorrefmark{2}Department of Electronic Engineering, Universitat Polyt\'ecnica de Catalunya, Barcelona, Spain}
\IEEEauthorblockA{\IEEEauthorrefmark{3}Department of Quantum and Computer Engineering, Delft University of Technology, Delft, The Netherlands}\vspace{-22pt}}

\maketitle

\begin{abstract}
Memristor networks are capable of low-power and massive parallel processing and information storage. Moreover, they have presented the ability to apply for a vast number of intelligent data analysis applications targeting mobile edge devices and low power computing. Beyond the memory and conventional computing architectures, memristors are widely studied in circuits aiming for increased intelligence that are suitable to tackle complex problems in a power and area efficient manner, offering viable solutions oftenly arriving also from the biological principles of living organisms. In this paper, a memristive circuit exploiting the dynamics of oscillating networks is utilized for the resolution of very popular and $NP$-complete logic puzzles, like the well-known ``Sudoku". More specifically, the proposed circuit design methodology allows for appropriate usage of interconnections' advantages in a oscillation network and of memristor's switching dynamics resulting to logic-solvable puzzle-instances. The reduced complexity of the proposed circuit and its increased scalability constitute its main advantage against previous approaches and the broadly presented SPICE based simulations provide a clear proof of concept of the aforementioned appealing characteristics.
\end{abstract}


\section{Introduction}
Conventional computers that are using CMOS technology in von Neumann architecture, are still the most dominant in today's computing, even if their scaling rate is foreseen to be soon saturated owing to the upcoming CMOS technology's physical limitations. Despite their efficiency in ``conventional" computations based on Boolean logic representation, they lack of performance against problems with increased complexity, which cannot be efficiently mapped to Boolean operations. Although there are plenty of existing algorithms endeavouring to solve this category of complex problems by utilising the ``conventional" computers, the scaling up of such problems is exponentially increasing the required time and/or computation resources to reach adequate solutions of these, namely Nondeterministic Polynomial time ($NP$-Complete) problems.

A special sub-category of the aforementioned complex problems are the logic puzzles, like the famous Japanese-origin  \begin{CJK}{UTF8}{min}S\={u}doku (数独)\end{CJK}\cite{delahaye2006science} called by the fusion of Japanese characters for ``number" (s\={u}) and ``single" (doku). This specific puzzle can be represented in various ways, like a constraint satisfaction problem or graph-coloring problem, and, consequently, different methods have been utilised towards its proper solution. However, the scaling of those solutions is mainly limited by the solving capabilities of the corresponding algorithms.

Alongside to those algorithms, the highly parallel processing capabilities of hardware accelerators, such Graphics Processing Units (GPUs), have been exploited for further reduction of the required solution time \cite{sato2011gpu}. Moreover, and beyond the software approaches, dedicated hardware has been also appropriately designed to increase the speed of computations against the widely used general purpose processors. Mostly, FPGA based implementations have been introduced that can effectively perform genetic and/or heuristic algorithms to solve Sudoku puzzles \cite{van2009tu,malakonakis2009fpga}.

Beyond conventional computers, novel electronic hardware approaches have been developed to provide scalable solution to those complex $NP$-Complete problems. Opposing to their conventional counterparts, those novel circuits and systems exploit the unprecedented characteristics of novel nano-devices, like Memristors \cite{chua2019handbook}, so as the solution emerges in an intrinsic manner from the dynamics of the designed approach. The novel memristor nano-scale devices are two-terminal passive circuit elements that can adapt their electrical conductance in response to the applied voltage or current on their terminals \cite{Chua71,chua2019handbook}. Memristors can also retain their conductance value as long as a voltage or current is not applied. The aforementioned novel characteristics for nano-device make memristor a competitive candidate for innovative memory and computing applications \cite{Tetzlaff2013,vourkasbook,chua2019handbook}.

Unconventional computing circuits that exploit memristor's nonlinearity and adaptivity to solve complex problems have been lately presented in literature \cite{adamatzky2011memristive,Alon2019}. Maze solving, shortest-path, bin packing, max-clique, graph-coloring, are just some of the complex $NP$-complete problems that memristors have been already utilized to provide a scalable solution, taking advantage from the computational topology of memristor networks and Cellular Automata (CA) \cite{stathis2014solving,Pershin13,Stathisiet,vourkasbook,parihar17,vourkas2015massively,Buscarino,Karamani}. Furthermore, memristor when combined with classic circuit topologies, like $RLC$ filters, to embody learning mechanisms in such simple circuitries \cite{ntinas2015lc}, has been successfully utilized to oscillation-based maze solving \cite{ntinas2017oscillation}. The problem-solving capabilities of the aforementioned oscillators with memristors have been efficiently tested through bio-inspired approaches that the behaviour and computational capabilities of microorganisms, such as Physarum Polycephalum \cite{Adamatzky2010,Pershin,Adamatzky2016} are modelled by the memristor-$RLC$ systems \cite{ntinas2017modeling}.

In this work, the complexity of memristor-$RLC$ oscillating circuits is reduced to simpler memristor-$RC$ circuits that can be easier integrated with less circuit elements and their oscillating dynamics have been exploited in puzzle solving related operations, namely Sudoku puzzle solving. Thus, a $3\times{3}$ Sudoku example representation and implementation are illustrated for reasons of readability and comprehensiveness, along with different example configurations and their corresponding solutions, produced by memristor-based oscillators. SPICE based simulations of the presented Sudoku configurations provide some insights referring to the appealing characteristics of the proposed memristor-$RC$ oscillating networks. 

\section{Oscillating Memristor-$RC$}
\label{Section:OScillation}
\begin{figure}[!t]
\centerline{\includegraphics[width=0.25\linewidth]{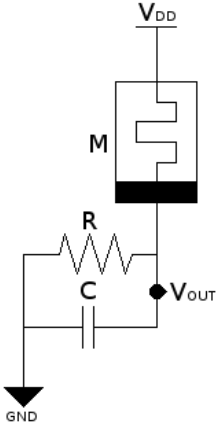}}
\caption{Memristor-RC oscillator.}
\label{fig:node}
\end{figure}

Taking advantage of the memristor's switching dynamics and the charging-discharging effects of the $RC$ circuit, the behaviour of a simplistic memristor-$RC$ oscillator is firstly discussed. Illustrated in Fig.~\ref{fig:node}, the oscillator consists of a memristor device in series with a $RC$ that are constantly excited by a \texttt{DC} source.

Aiming to preserve the circuit complexity of the oscillator in low levels, the circuit is powered by a single polarity constant voltage ($V_{DD}$), instead of the complementary way where both polarities are applied ($\pm{}V_{DD}$) and more circuitry is required. In other words, the switching operation of memristor devices has to be performed under the same voltage polarity. There are memristor devices, i.e. unipolar memristors, that both \texttt{SET} and \texttt{RESET} processes of their memristor's switching are triggered by the same polarity of the applied voltage. As a result, the proposed circuit has been designed with a unipolar Metal-Insulator-Semiconductor (MIS) device, whose dynamics are described in SPICE-compatible way by Redondo \textit{et al.} \cite{garcia2016spice}. The \texttt{SET} process of those devices is triggered by high amplitude applied voltage that effectively construct a Conductive Filament (\texttt{CF}) inside the insulating layer, achieving the Low Resistance State ($LRS$) of the devices. On the other hand, after the \texttt{CF}'s construction, a lower amplitude voltage can heat-up the \texttt{CF} and as a result of the Joule Heating, the most thin/sensitive part of the CF is disrupted and the device is driven back to the High Resistance State ($HRS$). In this specific unipolar memristor model, the switching of the device is attributed to the power dissipated on the device, where a power threshold for both \texttt{SET} and \texttt{RESET} processes has to be exceeded in order to trigger the correlated switching process.

\begin{figure}[!t]
\centerline{\includegraphics{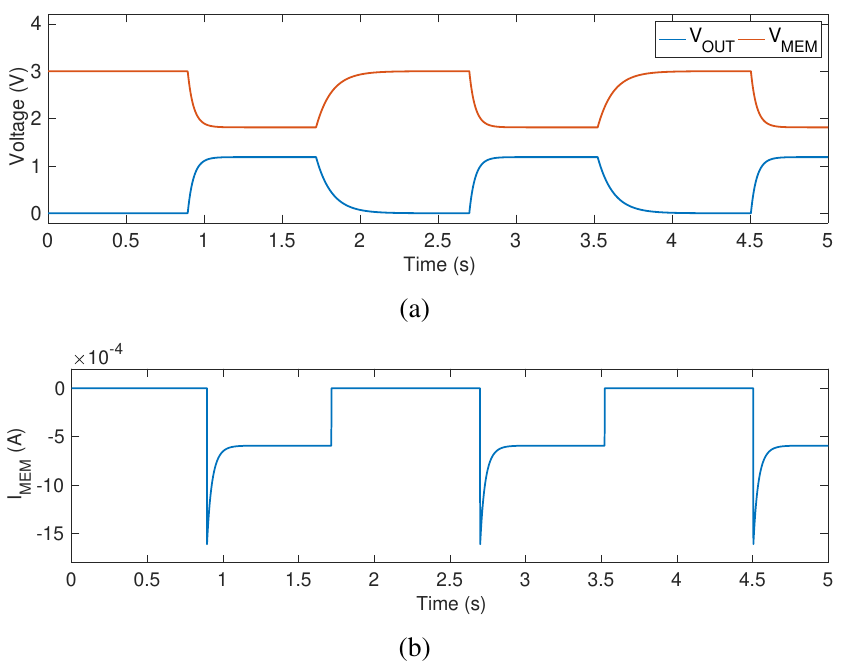}}
\caption{(a) Voltage at the Output of the node ($V_{OUT}$) and across memristor's terminals ($V_{MEM}$) over time. (b) Current ($I_{MEM}$) through the memristor over time.}
\label{fig:node_V-I}
\end{figure}

In the view of the aforementioned, in the design of the memristor-$RC$ oscillator, the values of the resistor $R$ and the capacitor $C$ have been chosen in an appropriate way that the oscillator should not reach an equilibrium point where no further memristor switching is performed. More specifically, the resistance value of $R$ defines the voltage at the Output node ($V_{OUT}$) of the oscillator, governed by the voltage divider between $R$ and Memristor's resistance $M$. In addition, the capacitance of $C$ controls efficiently the time required for the voltage to reach the voltage divider value at $V_{OUT}$. As a result, when the memristance $M$ is found in $HRS$, the most of the \texttt{DC} source voltage is applied on the memristor, as a result of the following design selection:

\begin{equation}
LRS<R\ll{}HRS
\label{eq1}
\end{equation}

\noindent and, consequently, the \texttt{CF} of memristor is constructed and $M$ is driven to $LRS$. Therefore, in $LRS$, the voltage at $V_{OUT}$ is increasing due to the voltage divider between $R$ and memristor's $LRS$, and, consequently, $C$ is charging. As long as $V_{OUT}$ is increasing, the applied voltage across memristor's terminals is decreasing as presented in Fig.~\ref{fig:node_V-I}; however, it is enough to heat-up the device and trigger its \texttt{RESET} process after some time. Then, the capacitor is discharging and the whole process of switching between $HRS$ and $LRS$ is iterated, generating a continuous oscillation of voltage at the $V_{OUT}$ node.

\begin{figure}[!t]
\centerline{\includegraphics[width=\linewidth]{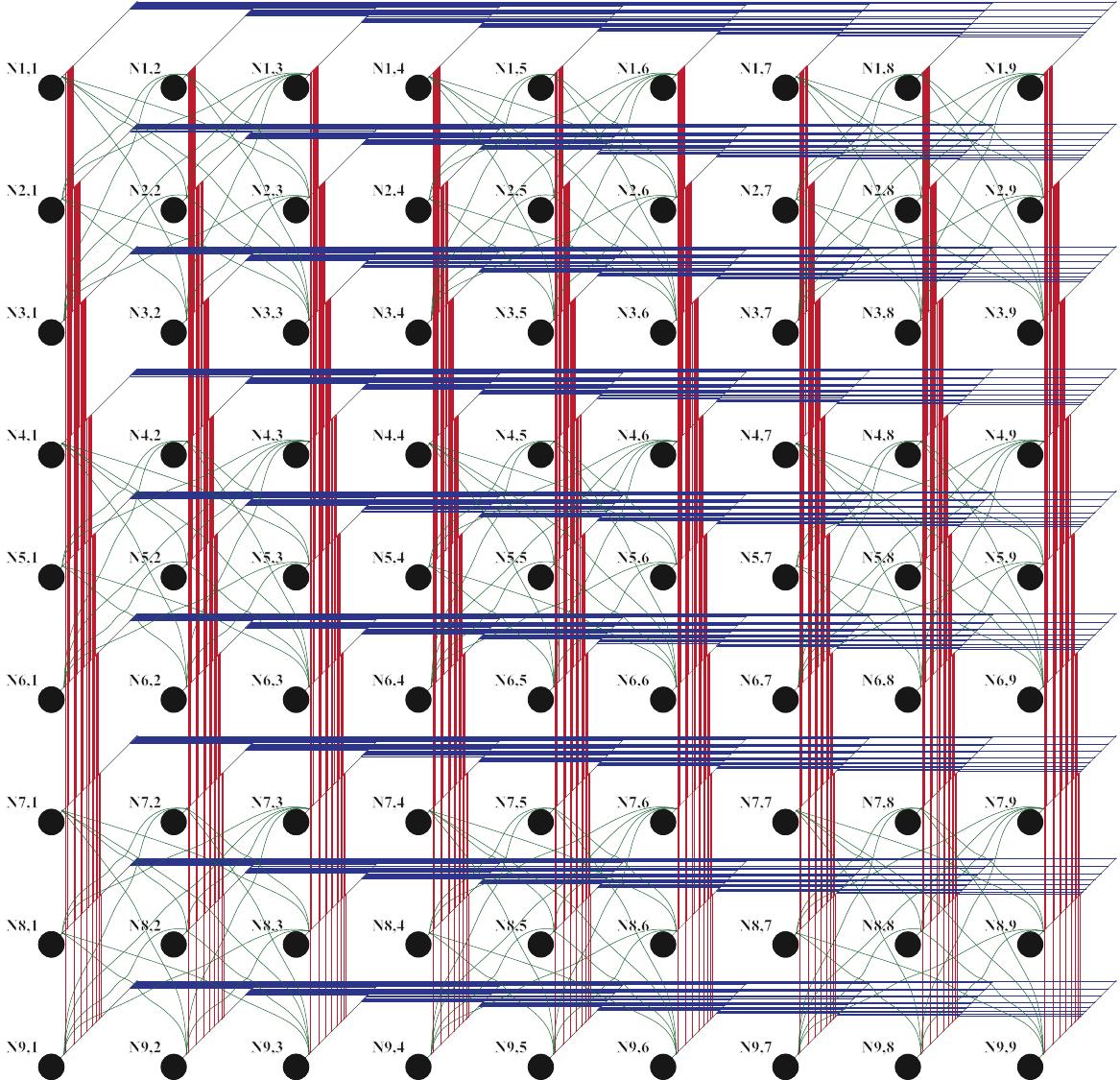}}
\caption{Sudoku cell interconnections for solving of the famous $9\times{}9$ Sudoku puzzle.}
\label{fig:9by9}
\end{figure}

\begin{figure*}[!t]
\centerline{\includegraphics{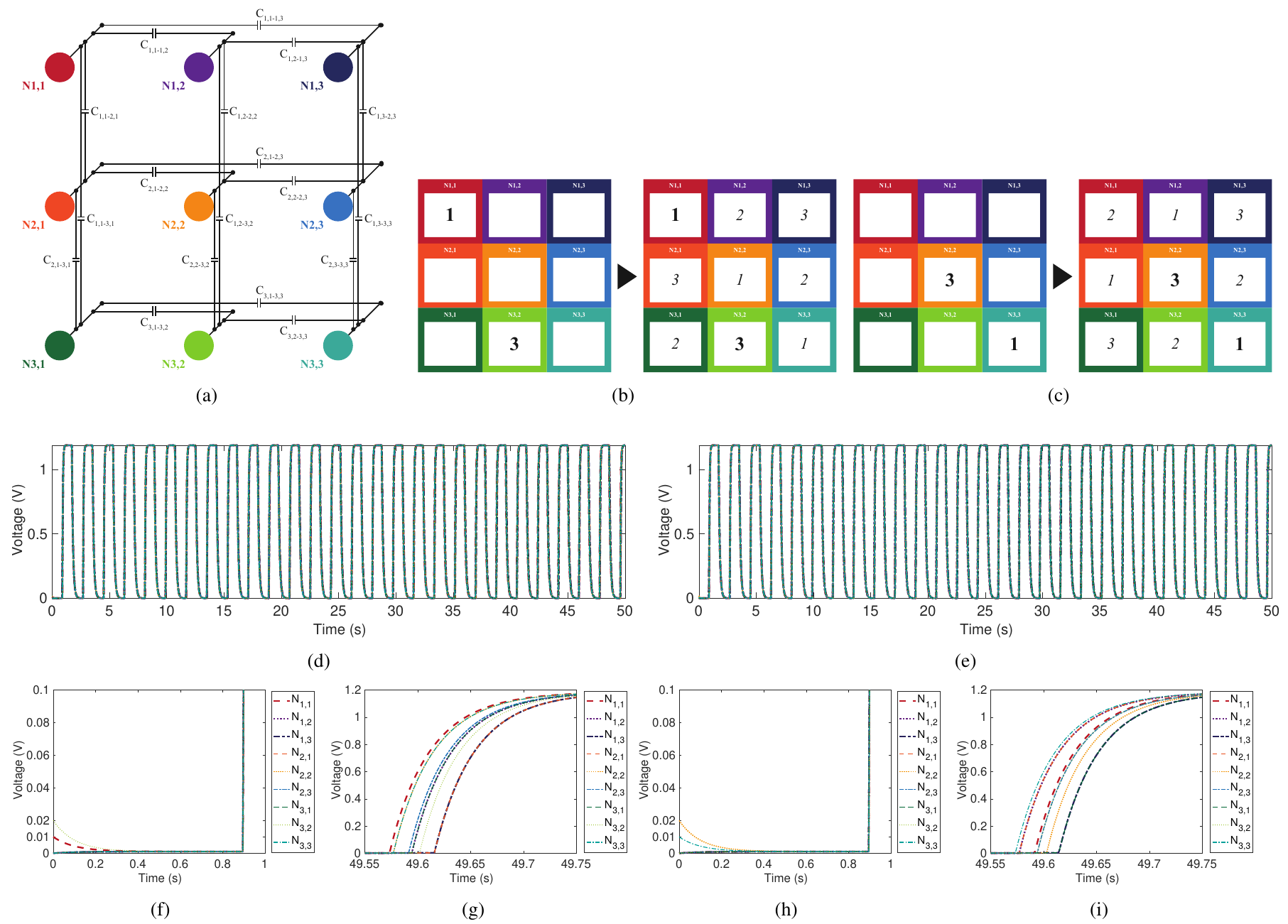}}
\caption{(a) $3\times{3}$ memristor-$RC$ oscillating network circuit for the representation of a $3\times{3}$ Sudoku puzzle, (b, c) Color-coded $3\times{}3$ Sudoku puzzle in two different configurations and their corresponding arithmetic solutions. (d, e) Output node voltage $V_{OUT}$ of $N_{i,j}$ for $i,j = 1, 2, 3$, representing the above different Sudoku configurations resulting to an appropriate solution. Zoom in (f, h) initialisation time and (g, i) last oscillation of the above Sudoku examples for the selected time. The lines in every plot follow the same color-coding as (a, b, c).}
\label{fig:3x3_res}
\end{figure*}

\section{Sudoku Representation and Solution Results}
\label{Section:Sudoku}
Having identified the basic principles of the memristor-$RC$ oscillating network, we are going to employ the presented dynamics for the solution of logic puzzles like Sudoku. More specifically, Sudoku as a mind-bending puzzle is proven to belong to the hard-to-be-solved type of computational problems from Yato and Seta \cite{yato2003complexity}. Its solution difficulty can be easily derived from the fact that Sudoku has a strong positive impact on adult’s brain and more specifically in memory performance as Grabbe implies in \cite{grabbe2017sudoku}. 

As it has been provided by Delahaye \cite{delahaye2006science}, Sudoku is a square grid puzzle that, typically, contains $n^2$ cells - divided into $n$ rows and $n$ columns -- and is grouped into $n$ smaller sub-grids with $n$ cells each. At the beginning, some of the cells are filled with pre-defined numbers and the solver has to fill the rest blank cells with any number of a given integer number set $K=\{1, 2, ..., n\}$. Nevertheless, there are solution constraints associated with the fact that the empty cells must be filled in such a way that the appearance of the same number in the any row, column or sub-grid is prohibited. As a result, well-conceived Sudoku puzzle must have one and only solution \cite{delahaye2006science}. The progenitor of Sudoku is the Latin square - a square matrix of $n^2$ cells - filled with $n$ symbols in such way the appearance of the same symbol in any row or column is prohibited. That leads to the use of each of the $n$ symbols exactly $n$ times in the square matrix. The classic version of Sudoku is a Latin square with dimensions $9\times{9}$ with the only difference being that each sub-grid must contain all the numbers from 1 through 9.

In order to map the Sudoku puzzle to an electronic circuitry that will automatically propose a solution, the memristor-$RC$ oscillator is used as a Sudoku cell ($N$) and a grid of $n\times{n}$ properly interconnected memristive-$RC$ oscillators constitutes the whole $n\times{n}$ Sudoku puzzle ($N_{i,j}$, for $i, j = 1, ..., n$). The oscillators are connected through coupling capacitors in a way that only the Sudoku cells, which are unable to have the same number, are connected directly. Hence, one oscillator's $V_{OUT}$ is directly connected through $\{C_{i, j} | i, j = 1, ..., n~\&~i\ne{}j\}$ with every other oscillator of its row, column and sub-group. An abstract representation of the proposed system for the classic $9\times{9}$ Sudoku puzzle is shown in Fig.~\ref{fig:9by9}.

Moreover, when the output voltage $V_{OUT}$ of an oscillator $N_{i,j}$ is increasing, results to the increase at the $V_{OUT}$ of a directly connected oscillator $N_{k,l}$. Subsequently, the voltage across the memristor of the latter oscillator $N_{k,l}$ is reduced so the switching of memristor's state is delayed, delaying also the operation of $N_{k,l}$. Utilising this delay, the proposed system is able to separate the directly connected oscillators, meaning that the nondirectly connected oscillators are slowly synchronised. As a result, the solution of the Sudoku puzzle is emerging through the timing separation of the oscillators, such as the cells with the same number of set $K$ will be oscillating together.

For the definition of the starting predefined numbers of a Sudoku example, the initial condition of capacitor $C_{i}$ of the corresponding oscillators is selected properly. By defining appropriate but different initial conditions for the capacitors, the initial voltage across the memristor of the corresponding oscillator is different, so that the memristor's switching time will be initially differentiated. 

Figs.~\ref{fig:3x3_res}(b, c) illustrate different examples of $3\times{3}$ Sudoku puzzles that are replicated and solved by the proposed memristive network circuitry shown in Fig. \ref{fig:3x3_res}(a) with both presented cases having two predefined numbers illustrated but with varying positions across the puzzle. At the left of each subfigure the initial $3\times{3}$ Sudoku example puzzle is presented and on the right side, the corresponding and unique possible solution is provided. During the mapping of these examples to the proposed circuit, the $V_{OUT}$ of all the nodes of these initial configurations of $3\times{3}$ Sudoku puzzles are represented in Figs.~\ref{fig:3x3_res}(d, e), respectively. For further clarification, Figs.~\ref{fig:3x3_res}(f, h) depict the initialization of the proposed system following the left Figs.~\ref{fig:3x3_res}(b, c), accordingly, where two nodes are initialised to different voltage values across their internal capacitors, representing the fixed numbers ``1" and ``3" of these examples placed in different positions. At the end of the simulations, the switching time of each oscillation has been adapted according to its connections. Figs.~\ref{fig:3x3_res}(g, i) focus on the beginning of the last increase of all oscillators, where the differentiation of switching time between the oscillators is evident. In both cases, the oscillators can be grouped into 3 discrete groups of 3 nodes, respectively, by their switching time from left to right in order to represent the 3 different numbers $K=\{1, 2, 3\}$ required by the presented Sudoku examples. It should be also mentioned that for sake of simplicity, the colours of the lines in Figs.~\ref{fig:3x3_res}(d-i) follow the exact color coding of the cells in Figs.~\ref{fig:3x3_res}(a, b, c).

Focusing on the presented simulation results, one can notice that for varying initial conditions as shown in Figs.~\ref{fig:3x3_res}(f, h), the oscillator's rising time has been separated in a different way in Figs.~\ref{fig:3x3_res}(g, i), reaching each time the different solution that is arithmetically presented in the corresponding Sudoku configuration Figs.~\ref{fig:3x3_res}(b, c).
\vspace{-12pt}
\section{Discussion}
The presented example provides a proof of concept for the proper functioning of the proposed Sudoku solving circuit. This specific example memristive circuitry of $3\times{3}$ Sudoku corresponding puzzles lacks the sub-group connections due to its small size; however, it has been selected for readability reasons as the illustration of large Sudoku results, as can be easily considered through the presented figures was found limited. Nevertheless, scaling-up the proposed circuit methodology and by designing a re-programmable array of interconnections between nodes will open a wide spectrum of potential applications in the area of computational problem solving and especially, in the area of logic puzzles solving. The solution is generated intrinsically from the dynamics of the designed system, which is also appropriate for large scale integration, so a future generic logic puzzle problem solver can be foreseen and designed taking into account the numerous variations of existing puzzles, like Sudoku. As presented by the simulation results, the initial conditions of the oscillators are essential for the convergence of the system to a solution. A further exploration of the system's behavior in the initial conditions domain is more than reasonable to reach a deeper understanding of its capabilities and enable, if possible, an automated selection of the initial states according to the requested application.
\vspace{-10pt}
\section{Conclusions}
In this paper, a memristor-based oscillator has been utilised to address the resolution of Sudoku puzzles. Despite the high complexity of such a task, the presented circuit is able to intrinsically come to a solution, without the use of complicated algorithmic methods. Moreover, the designed circuit comprises only of nano-scale integration compatible electronics elements, as it utilises memristor's nonlinear dynamics along with a classic $RC$ circuit for the generation of oscillations. As a proof of concept, in this work, an example of $3\times{3}$ Sudoku puzzle is solved for two different configurations, only by changing the oscillator's initial conditions.
\vspace{-12pt}
\bibliographystyle{IEEEtran}

\begin{thebibliography}{10}
\bibitem{delahaye2006science}
J.-P. Delahaye, ``The science behind {S}udoku,'' \emph{Scientific American},
  vol. 294, no.~6, pp. 80--87, 2006.

\bibitem{sato2011gpu}
Y.~Sato, N.~Hasegawa, and M.~Sato, ``{GPU} acceleration for {S}udoku solution
  with genetic operations,'' in \emph{2011 IEEE Congress of Evolutionary
  Computation (CEC)}.\hskip 1em plus 0.5em minus 0.4em\relax IEEE, 2011, pp.
  296--303.

\bibitem{van2009tu}
K.~Van Der~Bok, M.~Taouil, P.~Afratis, and I.~Sourdis, ``The {TU} {D}elft
  sudoku solver on {FPGA},'' in \emph{2009 International Conference on
  Field-Programmable Technology}.\hskip 1em plus 0.5em minus 0.4em\relax IEEE,
  2009, pp. 526--529.

\bibitem{malakonakis2009fpga}
P.~Malakonakis, M.~Smerdis, E.~Sotiriades, and A.~Dollas, ``An {FPGA}-based
  sudoku solver based on simulated annealing methods,'' in \emph{2009
  International Conference on Field-Programmable Technology}.\hskip 1em plus
  0.5em minus 0.4em\relax IEEE, 2009, pp. 522--525.

\bibitem{chua2019handbook}
L.~Chua, G.~C. Sirakoulis, and A.~Adamatzky, \emph{Handbook of Memristor
  Networks}.\hskip 1em plus 0.5em minus 0.4em\relax Springer Science \&
  Business Media, 2019.

\bibitem{Chua71}
L.~{Chua}, ``Memristor-the missing circuit element,'' \emph{IEEE Transactions
  on Circuit Theory}, vol.~18, no.~5, pp. 507--519, Sep. 1971.

\bibitem{Tetzlaff2013}
R.~Tetzlaff, \emph{Memristors and Memristive Systems}, ser. SpringerLink :
  B{\"u}cher.\hskip 1em plus 0.5em minus 0.4em\relax Springer New York, 2013.

\bibitem{vourkasbook}
I.~Vourkas and G.~C. Sirakoulis, \emph{Memristor-Based Nanoelectronic Computing
  Circuits and Architectures}.\hskip 1em plus 0.5em minus 0.4em\relax Springer
  International Publishing, 2016.

\bibitem{adamatzky2011memristive}
A.~Adamatzky and L.~Chua, ``Memristive excitable cellular automata,''
  \emph{International Journal of Bifurcation and Chaos}, vol.~21, no.~11, pp.
  3083--3102, 2011.

\bibitem{Alon2019}
R.~{Tetzlaff}, A.~{Ascoli}, I.~{Messaris}, and L.~O. {Chua}, ``Theoretical
  foundations of memristor cellular nonlinear networks: Memcomputing with
  bistable-like memristors,'' \emph{IEEE Transactions on Circuits and Systems
  I: Regular Papers}, pp. 1--14, 2019.

\bibitem{stathis2014solving}
D.~Stathis, I.~Vourkas, and G.~C. Sirakoulis, ``Solving {AI} problems with
  memristors: {A} case study for optimal,'' in \emph{Proceedings of the 18th
  Panhellenic Conference on Informatics}.\hskip 1em plus 0.5em minus
  0.4em\relax ACM, 2014, pp. 1--6.

\bibitem{Pershin13}
Y.~V. Pershin and M.~Di~Ventra, ``Self-organization and solution of
  shortest-path optimization problems with memristive networks,'' \emph{Phys.
  Rev. E}, vol.~88, p. 013305, Jul 2013.

\bibitem{Stathisiet}
I.~Vourkas, D.~Stathis, and G.~C. Sirakoulis, ``Memristor-based parallel
  sorting approach using one-dimensional cellular automata,'' \emph{Electronics
  Letters}, vol.~50, no.~24, pp. 1819--1821, November 2014.

\bibitem{parihar17}
A.~Parihar, N.~Shukla, M.~Jerry, S.~Datta, and A.~Raychowdhury, ``Vertex
  coloring of graphs via phase dynamics of coupled oscillatory networks,''
  \emph{Scientific Reports}, vol.~7, no.~1, p. 911, 2017.

\bibitem{vourkas2015massively}
I.~Vourkas, D.~Stathis, and G.~C. Sirakoulis, ``Massively parallel analog
  computing: {A}riadne’s thread was made of memristors,'' \emph{IEEE
  Transactions on Emerging Topics in Computing}, vol.~6, no.~1, pp. 145--155,
  2015.

\bibitem{Buscarino}
A.~{Buscarino}, C.~{Corradino}, L.~{Fortuna}, M.~{Frasca}, and L.~O. {Chua},
  ``Turing patterns in memristive cellular nonlinear networks,'' \emph{IEEE
  Transactions on Circuits and Systems I: Regular Papers}, vol.~63, no.~8, pp.
  1222--1230, Aug 2016.

\bibitem{Karamani}
R.~{Karamani}, I.~{Fyrigos}, V.~{Ntinas}, I.~{Vourkas}, and G.~C. {Sirakoulis},
  ``Game of life in memristor cellular automata grid,'' in \emph{CNNA 2018; The
  16th International Workshop on Cellular Nanoscale Networks and their
  Applications}, Aug 2018, pp. 1--4.

\bibitem{ntinas2015lc}
V.~Ntinas, I.~Vourkas, and G.~C. Sirakoulis, ``{LC} filters with enhanced
  memristive damping,'' in \emph{2015 IEEE International Symposium on Circuits
  and Systems (ISCAS)}.\hskip 1em plus 0.5em minus 0.4em\relax IEEE, 2015, pp.
  2664--2667.

\bibitem{ntinas2017oscillation}
V.~Ntinas, I.~Vourkas, G.~C. Sirakoulis, and A.~I. Adamatzky,
  ``Oscillation-based slime mould electronic circuit model for maze-solving
  computations,'' \emph{IEEE Transactions on Circuits and Systems I: Regular
  Papers}, vol.~64, no.~6, pp. 1552--1563, 2017.

\bibitem{Adamatzky2010}
A.~Adamatzky, \emph{Physarum Machines: Computers from Slime Mould}, ser. World
  Scientific series on nonlinear science.\hskip 1em plus 0.5em minus
  0.4em\relax World Scientific, 2010.

\bibitem{Pershin}
Y.~V. Pershin, S.~La~Fontaine, and M.~Di~Ventra, ``Memristive model of amoeba
  learning,'' \emph{Phys. Rev. E}, vol.~80, p. 021926, Aug 2009.

\bibitem{Adamatzky2016}
A.~Adamatzky, \emph{Advances in Physarum Machines: Sensing and Computing with
  Slime Mould}, ser. Emergence, Complexity and Computation.\hskip 1em plus
  0.5em minus 0.4em\relax Springer International Publishing, 2016.

\bibitem{ntinas2017modeling}
V.~Ntinas, I.~Vourkas, G.~Sirakoulis, and A.~Adamatzky, ``Modeling physarum
  space exploration using memristors,'' \emph{Journal of Physics D: Applied
  Physics}, vol.~50, no.~17, p. 174004, 2017.

\bibitem{garcia2016spice}
F.~Garcia-Redondo, R.~P. Gowers, A.~Crespo-Yepes, M.~L{\'o}pez-Vallejo, and
  L.~Jiang, ``{SPICE} compact modeling of bipolar/unipolar memristor switching
  governed by electrical thresholds,'' \emph{IEEE Transactions on Circuits and
  Systems I: Regular Papers}, vol.~63, no.~8, pp. 1255--1264, 2016.

\bibitem{yato2003complexity}
T.~Yato and T.~Seta, ``Complexity and completeness of finding another solution
  and its application to puzzles,'' \emph{IEICE Transactions on Fundamentals of
  Electronics, Communications and Computer Sciences}, vol.~86, no.~5, pp.
  1052--1060, 2003.

\bibitem{grabbe2017sudoku}
J.~W. Grabbe, ``Sudoku and changes in working memory performance for older
  adults and younger adults,'' \emph{Activities, Adaptation \& Aging}, vol.~41,
  no.~1, pp. 14--21, 2017.

\end{thebibliography}

\end{document}